\documentstyle[12pt]{article}
 \newcommand \be {\begin{equation}}
\newcommand \bea {\begin{eqnarray} \nonumber }
\newcommand \ee {\end{equation}}
\newcommand \eea {\end{eqnarray}}

 \newcommand \lan {\langle}
 \newcommand \ran {\rangle}

\newcommand{\bit}{\begin{itemize}}
\newcommand{\eit}{\end{itemize}}
\def\i{{\mathrm i}}

\begin{document}
\topskip 2cm 
\begin{titlepage}

\begin{center}
{\large\bf MIXING TRANSFORMATIONS IN QUANTUM FIELD THEORY} \\
\vspace{.3cm}
{\large\bf AND NEUTRINO OSCILLATIONS} \\
\vspace{2.5cm}
{\large M.Blasone$^{\dagger *}$\footnote{blasone@vaxsa.csied.unisa.it}, 
P.A.Henning$^{*}$\footnote{phenning@tpri6c.gsi.de}
and G.Vitiello$^{\dagger}$\footnote{vitiello@vaxsa.csied.unisa.it \\ \\
To appear in Proceedings of "Results and Perspectives in Particle Physics",
La Thuile, Aosta Valley, March 1996}} \\
\vspace{.5cm}
{\sl ${}^{\dagger}$Dipartimento di Fisica dell'Universit\`a}  \\       
{\sl and INFN, Gruppo Collegato, Salerno} \\
{\sl I-84100 Salerno, Italy. }\\
\vspace{.5cm}
{\sl ${}^{*}$Theoretical Physics,
        Gesellschaft f\"ur Schwerionenforschung GSI} \\              
{\sl P.O.Box 110552, D-64220 Darmstadt, Germany} \\ 

\vspace{2.5cm}

\begin{abstract}
Field mixing transformations are studied in quantum field theory
and the generator of the transformations is found to induce an SU(2)
coherent structure in the vacuum state, both for bosons and for fermions.
The Fock space for mixed fields
is unitarily inequivalent to the Fock space of the
unmixed fields in the infinite volume limit. We study
neutrino mixing and oscillations and find that the oscillation amplitude 
is depressed by a factor which is momentum and mass dependent. 
The usual formula is recovered in the
relativistic limit. Phenomenological features of the modified oscillation 
formula are discussed. Finally, preliminary results of the
Green's function formalism are presented.
\end{abstract}

\end{center}
\end{titlepage}

\newpage

\section{Introduction}

We report about recent results in the study of mixing transformations in 
quantum field theory (QFT) \cite{aop, pl}, with special attention to the 
case of neutrino mixing. There are indeed unexpected features in 
field mixing transformations which, as we will show below, find their 
origin in the same structure of QFT, in particular in the existence in QFT 
of infinitely many inequivalent
representations of the canonical commutation relations [3,4]. We 
find that the generator of mixing transformations induces a non
trivial structure in the vacuum which turns out to be a coherent
state, both for bosons and for fermions. The Fock space for mixed fields
appears to be unitarily inequivalent 
to the Fock space of the original (unmixed) fields in the 
infinite volume limit.

The question arises if such a new
structure leads to any possibly testable effect. For such a 
purpose, we investigate in detail
neutrino mixing and oscillations [5,6] and
a new
oscillation formula, different from the usual one, is obtained.
A correction on the oscillation amplitude is found which turns out to be 
momentum and mass dependent. In the
relativistic limit the usual formula is recovered; this is in general
agreement with other studies of neutrino oscillations in the non
relativistic region [7].

In Sec.2 we consider boson field mixing as well as fermion field mixing. 
In Sec.3 we study the neutrino mixing and oscillations and comment upon
phenomenological implications. In Sec.4 we present preliminary 
results on the Green's functions approach to mass mixing.

\section{Mixing transformations in quantum field theory}

Let us first discuss the case of charged boson mixing.
Consider two charged boson fields $\phi_{i}(x)$, $i=1,2$ and their 
conjugate momenta $\pi_{i}(x)=\partial_{0}\phi_{i}^{\dag}(x)$, satisfying 
the usual commutation relations with non-zero commutators given by:
\be 
\left[\phi_{i}(x),\pi_{i}(y)\right]_{t=t'}=
\left[\phi_{i}^{\dag}(x),\pi_{i}^{\dag}(y)\right]_{t=t'}=i\delta^{3}(\bar{x}-
\bar{y})
\;,\;\;\;\;\;\;
\left[a_{k,i},a_{p,i}^{\dag} \right]=
\left[b_{k,i},b_{p,i}^{\dag} \right]=\delta^{3}(\bar{k}-\bar{p})
\ee
\be
\phi_{i}(x) = \int  
\frac{d^{3}\bar{k}}{(2\pi)^{3 \over 2}} \frac{1}{\sqrt{2\omega_{i}}}
\left( a_{k,i}\:e^{-i k.x} + b^{\dag }_{k,i}\: 
e^{i k.x} \right)
\ee
\be
\pi_{i}(x) = \int 
\frac{d^{3}\bar{k}}{(2\pi)^{3 \over 2}} \sqrt{\frac{\omega_{i}}{2}}
\;i\left( a^{\dag }_{k,i}\:e^{i k.x} - b_{k,i}\: 
e^{-i k.x} \right)
\ee
with $k.x= \omega t - \bar{k} \cdot \bar{x}$. Here and in the following
we omit vector notation in the indices. The mixing relations are: 
\bea
\phi_{A}(x) = \phi_{1}(x) \; \cos\theta   + \phi_{2}(x) \; \sin\theta  
  \\ \phi_{B}(x) =- \phi_{1}(x) \; \sin\theta   + \phi_{2}(x)\; \cos\theta
\eea
and h.c. and similar ones for $\pi_{A}$ and $\pi_{B}$. 
We put them into the form:
\bea
\phi_{A}(x) = G^{-1}(\theta,t)\; \phi_{1}(x)\; G(\theta,t) \\ 
\phi_{B}(x) = G^{-1}(\theta,t)\; \phi_{2}(x)\; G(\theta,t)
\eea
and similar ones for $\pi_{A}$ and $\pi_{B}$; $G(\theta,t)$ is given by
\be 
G(\theta,t) = \exp\left[-i\;\theta \int d^{3}\bar{x} 
\left(\pi_{1}(x)\phi_{2}(x) - \pi_{2}^{\dag}(x)\phi_{1}^{\dag}(x)
-\pi_{2}(x)\phi_{1}(x) + \pi_{1}^{\dag}(x)\phi_{2}^{\dag}(x)\right)\right]
\ee
and is (at finite volume) an unitary operator:
$G^{-1}(\theta,t)=G(-\theta,t)=G^{\dag}(\theta,t)$.
By introducing the operator 
$S_{+} \equiv -i\;\int d^{3}\bar{x} \; (\pi_{1}(x)\phi_{2}(x) 
- \pi_{2}^{\dag}(x)\phi_{1}^{\dag}(x))$  (and $S_{-} =
\left( S_{+}\right)^{\dag}$),
$G(\theta,t)$ can be written as
\be
G(\theta,t) = \exp[\theta(S_{+} - S_{-})] ~.
\ee
It is easy to verify that, introducing $S_{3}$ and the total charge
$S_{0}$ as follows
\be
 S_{3} \equiv \frac{-i}{2} \int d^{3}\bar{x} 
\left(\pi_{1}(x)\phi_{1}(x) - \pi_{2}(x)\phi_{2}(x)
+\pi_{2}^{\dag}(x)\phi_{2}^{\dag}(x) 
- \pi_{1}^{\dag}(x)\phi_{1}^{\dag}(x)\right)
\ee
\be
 S_{0} ={Q \over 2}\equiv \frac{-i}{2} \int d^{3}\bar{x} 
\left( \pi_{1}(x)\phi_{1}(x) -\pi_{1}^{\dag}(x)\phi_{1}^{\dag}(x)  
+\pi_{2}(x)\phi_{2}(x) - \pi_{2}^{\dag}(x)\phi_{2}^{\dag}(x)
\right)
\ee
the $su(2)$ algebra is closed: $
 [S_{+} , S_{-}]=2S_{3} \;\;\;,\;\;\; [S_{3} , S_{\pm} ] = \pm S_{\pm}
\;\;\;,\;\;\;[S_{0} , S_{3}]= [S_{0} , S_{\pm} ] = 0$.
We can expand $S_{+}$ and $S_{-}$ as follows:
\be
S_{+}=\int d^{3}\bar{k} \left( U_{k}(t) \; a_{k,1}^{\dag}a_{k,2} - 
V_{k}^{*}(t) \; b_{-k,1}a_{k,2} + V_{k}(t) \; a_{k,1}^{\dag}b_{-k,2}^{\dag}
- U_{k}^{*}(t) \; b_{-k,1}b_{-k,2}^{\dag} \right)
\ee
\be
S_{-}=\int d^{3}\bar{k} \left( U_{k}^{*}(t) \; a_{k,2}^{\dag}a_{k,1} 
- V_{k}(t) \; a_{k,2}^{\dag}b_{-k,1}^{\dag}
+ V_{k}^{*}(t) \; b_{-k,2}a_{k,1} 
- U_{k}(t) \; b_{-k,2}b_{-k,1}^{\dag} \right)
\ee
with
$U_{k}(t)\equiv |U_{k}|\;e^{i(\omega_{1}-\omega_{2})t}$ ~, 
$V_{k}(t)\equiv |V_{k}|\;e^{i(\omega_{1}+\omega_{2})t}$
and
\be
|U_{k}|\equiv {1 \over 2} \left( \sqrt{\omega_{1} \over \omega_{2}}
+\sqrt{\omega_{2} \over \omega_{1}}  \right) ~,
\;\;\;\;\;\;
|V_{k}|\equiv {1 \over 2} \left( \sqrt{\omega_{1} \over \omega_{2}}
-\sqrt{\omega_{2} \over \omega_{1}}  \right) ~.
\ee
Since $|U_{k}|^{2}-|V_{k}|^{2}=1 $, one can put $|U_{k}|\equiv \cosh 
\sigma_{k}$ , $|V_{k}|\equiv \sinh \sigma_{k}$ with $\sigma_{k}= {1 \over 
2} \ln\left( {\omega_{1} \over \omega_{2}}\right)$.

Our main observation is now that {\it the generator of
mixing transformations does not leave invariant the vacuum of the
fields $\phi_{1,2}(x)$, say $|0 \rangle_{1,2}$, since it induces an $SU(2)$ 
coherent state structure in this state}
[8,1].
This coherent state is the vacuum for the fields 
$\phi_{A, B}(x)$, which we
denote by $|0 \rangle_{A,B}$:
\be
|0 \rangle_{A,B} = G^{-1}(\theta,t)\; |0 \rangle_{1,2} 
\ee
The annihilator operators for $|0 \rangle_{A,B}$ are 
given by $a_{k,A} \equiv G^{-1}(\theta,t) \; a_{k,1}\;G(\theta,t)$, etc.:
\be
a_{k,A}(t)=\cos\theta\;a_{k,1}\;+\;\sin\theta\;\left(
U_{k}(t)\; a_{k,2}\;+\; V_{k}(t)\; b^{\dag}_{-k,2}\right) ~.
\ee
Similar expressions can be obtained for $a_{k,B}$, $b_{k,A}$ and
$b_{k,B}$. We also have
\be
\;_{1,2}\lan 0| a_{k,A}^{\dag} a_{k,A} |0\ran_{1,2}= \sin^{2}\theta\;
|V_{k}|^{2} =\sin^{2}\theta\; \sinh^{2}\left[ 
{1 \over 2} \ln\left( {\omega_{1} \over \omega_{2}}\right)\right] ~.
\ee

Corresponding results are obtained in the case of two neutral boson fields 
$\phi_{i}(x)$, $i=1,2$ and their conjugate momenta 
$\pi_{i}(x)=\partial_{0}\phi_{i}(x)$, satisfying the usual commutation 
relations. The non-zero commutators are:
\be 
\left[ \phi_{i}(x),\pi_{i}(y)\right]_{t=t'}=
 i\delta^{3}(\bar{x}-\bar{y}) \;,\;\;\;\;\;
\left[ a_{k,i}, a_{p,i}^{\dag} \right]=
\delta^{3}(\bar{k}-\bar{p})
\ee
\be
\phi_{i}(x) = \int  
\frac{d^{3}\bar{k}}{(2\pi)^{3 \over 2}} \frac{1}{\sqrt{2\omega_{i}}}
\left( a_{k,i}\:e^{-i k.x} + a^{\dag }_{k,i}\: 
e^{i k.x} \right)
\ee
\be
\pi_{i}(x) = \int 
\frac{d^{3}\bar{k}}{(2\pi)^{3 \over 2}} \sqrt{\frac{\omega_{i}}{2}}
\;i\left(- a_{k,i}\:e^{-i k.x} + a^{\dag }_{k,i}\: 
e^{i k.x} \right)
\ee
with $k.x= \omega t - \bar{k} \cdot \bar{x}$. The generator
of the 
mixing relations (corresponding to eqs. (4-5)) now is given by
$G(\theta,t) = \exp\left[-i\;\theta \int d^{3}\bar{x} 
\left(\pi_{1}(x)\phi_{2}(x) -\pi_{2}(x)\phi_{1}(x) \right)\right]
$,
and again is (at finite volume) an unitary operator. By
introducing the operators $
 S_{+} \equiv -i\;\int d^{3}\bar{x} \; \pi_{1}(x)\phi_{2}(x) 
$ and $S_{-} \equiv -i\;\int d^{3}\bar{x} \;\pi_{2}(x)\phi_{1}(x)$,
$G(\theta,t)$ can be written as $G(\theta,t) = \exp[\theta(S_{+} - S_{-})]$.
Again, by introducing $S_{3}$ and the total "charge"
$S_{0}$ as follows
\be
 S_{3} \equiv \frac{-i}{2} \int d^{3}\bar{x} 
\left(\pi_{1}(x)\phi_{1}(x) - \pi_{2}(x)\phi_{2}(x)\right) \;, \;\;
S_{0} \equiv \frac{-i}{2} \int d^{3}\bar{x} \left( \pi_{1}(x)\phi_{1}(x) + 
\pi_{2}(x)\phi_{2}(x) \right)
\ee
the $su(2)$ algebra is closed  and by
expanding the $S$'s operators in terms of creation and annihilation 
operators expressions similar to eqs.(12-15) are obtained.

For fermion field mixing we consider directly the case of neutrino 
field for sake of shortness. However, in this 
section, our considerations apply to any Dirac field.
The two flavor mixing relations (for the case of three
flavors see ref.[1]) are:
\bea
\nu_{e}(x) = \nu_{1}(x) \; \cos\theta   + \nu_{2}(x) \;
 \sin\theta \\
\label{mix3}
\nu_{\mu}(x) =- \nu_{1}(x) \; \sin\theta   + \nu_{2}(x)\; 
\cos\theta\;.  
\eea
Here
$\nu_{e}(x)$ and $\nu_{\mu}(x)$ are the (Dirac) neutrino fields with 
definite flavors.
 $\nu_{1}(x)$ and $\nu_{2}(x)$ are the (free) 
neutrino fields with definite masses $m_{1}$ and $m_{2}$, respectively
(we do not need to distinguish between left-handed and right-handed
components):
\be
\nu_{i}(x) = \frac{1}{\sqrt{V}} \sum_{ k,r}[u^{r}_{k,i}(t)
\alpha ^{r}_{k,i}\:e^{i \bar{k} \cdot \bar{x}}+ v^{r}_{k,i}(t)
\beta ^{r\dag }_{k,i}\: e^{-i \bar{k} \cdot \bar{x}}], \; ~ i=1,2 \; ,
\ee
with 
$\alpha ^{r}_{k,i}|0\rangle_{12}= \beta ^{r }_{k,i}|0\rangle_{12}=0$,
$  i=1,2 \;, \;r=1,2$.
For simplicity, the vector $\bar{k}$ and
its modulus are denoted by the same symbol.
The non-zero anticommutation relations are:
\be
\{\nu^{\alpha}_{i}(x), \nu^{\beta\dag }_{j}(y)\}_{t=t'} = 
\delta^{3}(\bar{x}-\bar{y})
\delta _{\alpha\beta} \delta_{ij} \;, \;\;\;\;\; \alpha,\beta=1,..,4 \;,
\ee
\be
\{\alpha ^{r}_{k,i}, \alpha ^{s\dag }_{q,j}\} = \delta
_{kq}\delta _{rs}\delta _{ij}   ;\qquad \{\beta ^{r}_{k,i}, \beta ^{s\dag
}_{q,j}\} = \delta _{kq} \delta _{rs}\delta _{ij},\;\;\;\; 
i,j=1,2\;. 
\ee
The orthonormality and completeness relations are the usual ones.
As in the boson case, we rewrite the mixing relations (1) in the form:
$\nu_{e}^{\alpha}(x) = G^{-1}(\theta,t)\; \nu_{1}^{\alpha}(x) 
G(\theta,t)$,~$\nu_{\mu}^{\alpha}(x) = G^{-1}(\theta,t)\; 
\nu_{2}^{\alpha}(x)\; G(\theta,t)\;$, 
and the generator is $G(\theta , t) = \exp[\theta(S_{+} - S_{-})]$, 
with
\be
S_{+} \equiv  \int d^{3}\bar{x} \; \nu_{1}^{\dag}(x)
\nu_{2}(x) \;\;,\;\;\;
S_{-} \equiv  \int d^{3}\bar{x} \; \nu_{2}^{\dag}(x)
\nu_{1}(x)\;= \;S_{+}^{\dag}\;. 
\ee
It is easy to see, by introducing $ S_{3} 
\equiv \frac{1}{2} \int d^{3}\bar{x} (\nu_{1}^{\dag}(x)\nu_{1}(x) - 
\nu_{2}^{\dag}(x)\nu_{2}(x)) $, that the $su(2)$ algebra is closed:
$ [S_{+} , S_{-}]=2S_{3} \;,\; [S_{3} , S_{\pm} ] = \pm S_{\pm}$.
 
As in the boson case, the main point[1] is that the above generator of
mixing transformations does not leave invariant the vacuum of the free
fields $\nu_{1,2}$, say $|0 \rangle_{1,2}$, since it induces an $SU(2)$ 
coherent state structure of neutrino-antineutrino pairs in this state.
This coherent state is the vacuum for the fields $\nu_{e,\mu}$, which we
denote by $|0 \rangle_{e,\mu}$:
\be
|0 \rangle_{e,\mu} = G^{-1}(\theta,t)\; |0 \rangle_{1,2}\;. 
\ee

In the infinite volume limit
$|0 \rangle_{e,\mu}$ is orthogonal to $|0 \rangle_{1,2}$ which
exhibits the non unitary character of the mixing transformations[1].

We can then construct the Fock space for the mixed field operators 
which are written as:
\be
\nu_{e}(x) = \frac{1}{\sqrt{V}} \sum_{k,r}\:e^{i \bar{k} \cdot \bar{x}}
 [u^{r}_{k,1}(t)
\alpha ^{r}_{k,e}(t)\: + v^{r}_{-k,1}(t)
\beta ^{r\dag }_{-k,e}(t)] 
\ee
\be
\nu_{\mu}(x) = \frac{1}{\sqrt{V}} \sum_{k,r}\:e^{i \bar{k} \cdot \bar{x}}
 [u^{r}_{k,2}(t)
\alpha ^{r}_{k,\mu}(t)\: + v^{r}_{-k,2}(t)
\beta ^{r\dag }_{-k,\mu}(t)] 
\ee
where the wave functions for the massive fields have been used [1,2] 
and (in the reference frame  $k=(0,0,|k|)$) the creation and
annihilation operators are given by:
\be \label{op1}
\alpha^{r}_{k,e}(t)=\cos\theta\;\alpha^{r}_{k,1}\;+\;\sin\theta\;\left(
U_{k}^{*}(t)\; \alpha^{r}_{k,2}\;+\;\epsilon^{r}\;
V_{k}(t)\; \beta^{r\dag}_{-k,2}\right) 
\ee
\be
\alpha^{r}_{k,\mu}(t)=\cos\theta\;\alpha^{r}_{k,2}\;-\;\sin\theta\;\left(
U_{k}(t)\; \alpha^{r}_{k,1}\;-\;\epsilon^{r}\;
V_{k}(t)\; \beta^{r\dag}_{-k,1}\right)  
\ee
\be
\beta^{r}_{-k,e}(t)=\cos\theta\;\beta^{r}_{-k,1}\;+\;\sin\theta\;\left(
U_{k}^{*}(t)\; \beta^{r}_{-k,2}\;-\;\epsilon^{r}\;
V_{k}(t)\; \alpha^{r\dag}_{k,2}\right) 
\ee
\be\label{op4}
\beta^{r}_{-k,\mu}(t)=\cos\theta\;\beta^{r}_{-k,2}\;-\;\sin\theta\;\left(
U_{k}(t)\; \beta^{r}_{-k,1}\;+\;\epsilon^{r}\;
V_{k}(t)\; \alpha^{r\dag}_{k,1}\right) 
\ee
with $\epsilon^{r}=(-1)^{r}$ and
\be
V_{k}(t)=|V_{k}|\;e^{i(\omega_{k,2}+\omega_{k,1})t}\;\;\;\;,\;\;\;\;
U_{k}(t)=|U_{k}|\;e^{i(\omega_{k,2}-\omega_{k,1})t} 
\ee
\be
|U_{k}|=\left(\frac{\omega_{k,1}+m_{1}}{2\omega_{k,1}}\right)^{\frac{1}{2}}
\left(\frac{\omega_{k,2}+m_{2}}{2\omega_{k,2}}\right)^{\frac{1}{2}}
\left(1+\frac{k^{2}}{(\omega_{k,1}+m_{1})(\omega_{k,2}+m_{2})}\right) 
\ee
\be
|V_{k}|=\left(\frac{\omega_{k,1}+m_{1}}{2\omega_{k,1}}\right)^{\frac{1}{2}}
\left(\frac{\omega_{k,2}+m_{2}}{2\omega_{k,2}}\right)^{\frac{1}{2}}
\left(\frac{k}{(\omega_{k,2}+m_{2})}-\frac{k}{(\omega_{k,1}+m_{1})}\right) 
\ee
\be
|U_{k}|^{2}+|V_{k}|^{2}=1 
\ee
\be
|V_{k}|^{2}=|V(k,m_{1}, m_{2})|^{2}=
\frac{k^{2}\left[(\omega_{k,2}+m_{2})-(\omega_{k,1}+m_{1})\right]^{2}}
{4\;\omega_{k,1}\omega_{k,2}(\omega_{k,1}+m_{1})(\omega_{k,2}+m_{2})}
\ee
where $\omega_{k,i}=\sqrt{k^{2}+m_{i}^{2}}$.
The number
operator $N_{\sigma_{l}}^{k,r}$ has vacuum expectation value:
\be
\;_{1,2}\langle0|\;N_{\sigma_{l}}^{k,r}\;|0\rangle_{1,2}\;=
\;\sin^{2}\theta\;|V_{k}|^{2} \;,\;\;\;\; \sigma=\alpha, \beta \;,\;\;\;\;
l=e,\mu ,
\ee
This last equation gives the condensation density of the vacuum
state $|0\rangle_{1,2}$
as a function of the mixing angle $\theta$, of the
masses $m_{1}$ and $m_{2}$ and of the momentum modulus $k$. Notice 
the difference with the usual approximation case where 
one puts
$|0\rangle_{e,\mu}=|0\rangle_{1,2}
\equiv|0\rangle$ and it is $\langle0|\;N_{\alpha_{e}}^{k,r}\;
|0\rangle\;=\;\langle0|\;N_{\alpha_{\mu}}^{k,r}\;
|0\rangle\;=0\;$.
Also note that 
$_{1,2}\langle0|\;N_{\sigma_{l}}^{k,r}\;|0\rangle_{1,2}$
plays the role of zero point contribution when considering
the energy contribution of
${\sigma_{l}}^{k,r}$ particles [1].

Let us close this section with the following remarks.
To be definite let us consider the fermion field case. We observe that the 
mixing relations
(20) relate the "free" hamiltonian $H_{1,2}$ (consider only mass terms) and 
$H_{e,\mu}$ [6] which includes also interacting terms:
\be
H_{1,2}=m_{1}\;\bar {\nu}_{1} \nu_{1} + m_{2}\;\bar {\nu}_{2} \nu_{2} 
\ee
\be
H_{e,\mu}=m_{ee}\; \bar {\nu}_{e} \nu_{e} + 
m_{\mu\mu}\;\bar {\nu}_{\mu} \nu_{\mu}+
m_{e\mu}\left(\bar {\nu}_{e} \nu_{\mu} + \bar {\nu}_{\mu} \nu_{e}\right) 
\ee
where $m_{ee}=m_{1}\cos^{2}\theta + m_{2} \sin^{2} \theta$, 
$m_{\mu\mu}=m_{1}\sin^{2}\theta + m_{2} \cos^{2} \theta$ 
and $m_{e\mu}=(m_{2}-m_{1})\sin\theta \cos \theta$.
In the LSZ formalism of QFT [3,4] observables are expressed in terms of 
asymptotic in- (or out-) fields (also called free or physical fields)
obtained by weak limit of Heisenberg or interacting 
fields for $t \rightarrow - (or +)
\infty$. The dynamics, i.e. the Lagrangian and the resulting field 
equations, is given in terms of the Heisenberg fields. The meaning of 
weak limit is to provide a realization of the basic dynamics in terms of 
the asymptotic fields. Since infinitely many representations of the 
canonical (anti-)
commutation relations exist in QFT [3,4]
the weak limit is however not unique.
As a consequence the realization of 
the basic dynamics in terms of the asymptotic fields is not unique and 
therefore, in order to avoid ambiguities
(unitarily inequivalent representations describe physically different 
phases), much care is needed in the study of the mapping among Heisenberg 
fields and free fields (generally known as dynamical
mapping or Haag expansion) [3,4].

For example, in theories with spontaneous symmetry
breaking the same set of Heisenberg field equations describes the
normal (symmetric) phase as well as the symmetry broken phase, according 
to the representation one chooses for the asymptotic fields.

Notice that in quantum mechanics no problem arises
with uniqueness of the asymptotic limit since finite volume 
systems are considered. In such a case in fact, due to the von Neumann 
theorem, representations of the canonical commutation 
relations are each other unitary equivalent.
However, in QFT where infinite number of degrees of freedom is 
considered  the von Neumann
theorem does not hold and much care is then required when considering any 
mapping among interacting and free fields [3,4]. We have seen in fact that,
in the case of field mixing,  
$|0 \rangle_{e,\mu}$ is orthogonal to $|0 \rangle_{1,2}$
in the infinite volume limit.
Field mixing relations,
which can be seen as
a mapping
among Heisenberg fields and free fields,
deserve thus a
careful analysis for reasons intrinsic to the QFT structure.

\section{The neutrino oscillation formula}

The neutrino oscillation formula is obtained by using the mixing mappings 
(\ref{op1}-\ref{op4}) [1]:
$$
\langle\alpha_{k,e}^{r}(t)|\;N_{\alpha_{e}}^{k,r}\;
|\alpha_{k,e}^{r}(t)\rangle\;= 
$$
\be\label{osc1}
=\;1 -\;\sin^{2}\theta\;|V_{k}|^{2}\;
-\;|U_{k}|^{2}\;\sin^{2}2\theta
\;\sin^{2}\left(\frac{\Delta\omega_{k}}{2}t\right)\;. 
\ee  

The number of $\alpha_{\mu}^{k,r}$ particles in the same state is
$$
\langle\alpha_{k,e}^{r}(t)|\;N_{\alpha_{\mu}}^{k,r}\;
|\alpha_{k,e}^{r}(t)\rangle\;=
$$
\be\label{osc2}
=\;|U_{k}|^{2}\;\sin^{2}2\theta
\;\sin^{2}\left(\frac{\Delta\omega_{k}}{2}t\right)+
\;\sin^{2}\theta\;|V_{k}|^{2}\;
\left(1\;-\;\sin^{2}\theta\;|V_{k}|^{2}\right)\;.
\ee 

The vacuum condensate contributes with the terms with $|V_{k}|^{2}$ and 
$|U_{k}|^{2}$ in (\ref{osc1}) and (\ref{osc2}).
We observe that
\be
\langle\alpha_{k,e}^{r}(t)|\;N_{\alpha_{e}}^{k,r}\;
|\alpha_{k,e}^{r}(t)\rangle + 
\langle\alpha_{k,e}^{r}(t)|\;N_{\alpha_{\mu}}^{k,r}
\;|\alpha_{k,e}^{r}(t)\rangle= %
\langle\alpha_{k,e}^{r}|\;N_{\alpha_{e}}^{k,r}\;
|\alpha_{k,e}^{r}\rangle + 
\langle\alpha_{k,e}^{r}|\;N_{\alpha_{\mu}}^{k,r}
\;|\alpha_{k,e}^{r}\rangle\;. 
\ee
where $|\alpha_{k,e}^{r}\rangle = |\alpha_{k,e}^{r}(t = 0)\rangle$, which
shows the conservation of the number $(N_{\alpha_{e}}^{k,r}\;
+ N_{\alpha_{\mu}}^{k,r})$ . The expectation value of this number in the 
state $|0\rangle_{1,2}$ is not zero due to the condensate contribution.
The (approximate) relations corresponding to eqs.(\ref{osc1}) and
(\ref{osc2}) in the conventional treatment are[5,6]:
\be\label{osc3}
\langle\alpha_{k,e}^{r}(t)|\;N_{\alpha_{e}}^{k,r}\;
|\alpha_{k,e}^{r}(t)\rangle\;=\;
1-\sin^{2}2\theta\;\sin^{2}\left(\frac{\Delta\omega_{k}}{2}t
\right)\;  
\ee
and
\be\label{osc4}
\langle\alpha_{k,e}^{r}(t)|\;N_{\alpha_{\mu}}^{k,r}\;
|\alpha_{k,e}^{r}(t)\rangle\;=\;
\sin^{2}2\theta\;\sin^{2}\left(\frac{\Delta\omega_{k}}{2}t
\right)~, 
\ee
respectively.
The conventional (approximate) results (\ref{osc3}) and (\ref{osc4}) are 
recovered when the condensate contributions are missing (in the $|V_{k}| 
\rightarrow 0$ limit).
In conclusion, in the QFT treatment we obtain 
corrections to the flavor oscillations which come from the condensate 
contributions.

The function $|V_{k}|^{2}$ has been studied in ref.[2] where the 
phenomenological implications of the results (\ref{osc1}) and (\ref{osc2})
 have also been 
discussed. Note that $|V_{k}|^{2}$  depends on $k$ only through its modulus 
and it is always in the interval $[0,\frac{1}{2}[$. It has a maximum for 
$k= \sqrt{m_{1}m_{2}}$ and
$|V_{k}|^{2}=0$  when $m_{1} = m_{2}$.
Also, $|V_{k}|^{2} \rightarrow 0$ when $k \rightarrow \infty$.

The corrections disappear in the infinite momentum or relativistic
limit $k >> \sqrt{m_{1} m_{2}}$ ($\sqrt{m_{1} m_{2}}$ is the
scale of the condensation density).
However, 
for finite $k$, the oscillation amplitude is depressed by a factor 
$|U_{k}|^{2}$: this "squeezing" factor ranges from $1$ to $\frac{1}{2}$ 
depending on $k$ and on the masses values and thus it may have not
negligible effects in experimental observations. The dependence of the 
flavor oscillation amplitude on the momentum could thus be tested.

To study the effects of the momentum dependence $|V_{k}|^{2}$ is 
written as
\be
|V_{k}|^{2} \equiv |V(p,a)|^{2}=
\frac{1}{2}\left(1-\frac{1}{\sqrt{1+a\left(\frac{p}{p^{2}+1}\right)^{2}}}
\right)  
\ee
\be
p=\frac{k}{\sqrt{m_{1} m_{2}}}\;\;\;\;\;,\;\;\;\;a=\frac{(\Delta m)^{2}}
{m_{1} m_{2}}\;\;\;,\;\;\;0\leq a < + \infty ~, 
\ee
where $\Delta m \equiv m_{2}-m_{1}$ (we take $m_{1}\leq m_{2}$).
At $p=1$, $|V(p,a)|^{2}$ reaches its maximum value $|V(1,a)|^{2}$,
which goes 
asymptotically to 1/2 when $a \rightarrow \infty$.

It is useful to calculate the value of $p$, say $ p_{\epsilon}$, at which 
the function $|V(p,a)|^{2}$ becomes a 
fraction $\epsilon$ of its maximum value
$V(1,a)$:
\be
p_{\epsilon}=\sqrt{-c+\sqrt{c^{2}-1}}\;\;\;\;,\;\;\;\;
c\equiv\frac{b^{2}(a+2)-2}{2(b^{2}-1)} \;\;\;\;,\;\;\;\;
b\equiv1-\epsilon\left(1-\frac{2}{\sqrt{a+4}}\right)~.
\ee

{}From Tab.I-III we see that the oscillation amplitudes of neutrinos of not 
very large momentum may have sensible squeezing factors.
We observe that large passive 
detectors include
neutrino momentum as
low as few hundreds of keV [5] and therefore
deviations from the usual oscillation formula may be expected
in these low momentum ranges.

An interesting case[2] occurs when one of the two masses, say
$m_{1}$, goes to zero. Then the maximum of the condensation
density occurs at $k \simeq 0$; however, since 
$a \rightarrow \infty$ when $m_{1} \rightarrow 0$, it is still possible
to have sensible effects at rather ``large'' momenta; $m_{2}$ should be 
large in order to provide appreciable corrections. The
situation is illustrated in Tab. III, where for the calculation we used 
$m_{1}=10^{- 10} eV$.

Since $|U_{k}|^{2}$ has a minimum at $k= \sqrt{m_{1}m_{2}}$ the dependence 
of oscillating
amplitude on the 
momentum, if experimentally tested, may provide
 an indication on 
neutrino masses. 

As we have seen, the vacuum condensate induces
the correction factor; the vacuum thus acts as a "momentum (or spectrum) 
analyzer" for oscillating neutrinos: neutrinos with 
$k\gg\sqrt{m_{1}m_{2}}$ have oscillation amplitude larger
than neutrinos with $k\simeq\sqrt{m_{1}m_{2}}$, 
due to vacuum structure. This "vacuum spectral analysis" effect may sum 
up to other effects (such as MSW effect [9] in matter) in depressing 
or enhancing neutrino oscillations (see  ref.[1] for a generalization 
of the above scheme to oscillations in matter).

In conclusion, probing the non relativistic momentum domain may provide new 
insights into neutrino physics.

\section{Green's functions for mixed fields}

In this section we present preliminary results [10] of the Green's function 
formalism for mixed fields in QFT. We will consider the case of (Dirac) 
fermion fields.
As in the previous sections we consider the mixing problem only for two 
fields.

Let us observe that the flavor operator  $\alpha_e$ 
has contributions from $\alpha_1$, $\alpha_2$ but {\em also\/}
from the anti-particle operator $\beta^{\dag}_2$ (and similarly for other 
flavor operators in eqs.(\ref{op1}-\ref{op4})).
This additional contribution is due to the fact that the spinor wave
functions for different masses are not orthogonal. In the more traditional
treatment of mixing the $\beta^{\dag}_2$ contribution is missed since the
non-orthogonality of the spinor wave functions is not considered.
An important point here is time dependence
of flavor operators in eqs.(\ref{op1}-\ref{op4}).
The transition amplitude from a flavor eigenstate created by 
$\alpha_{k,e}^{r \dag}(t^\prime)$ at time $t^\prime$ into the same
state at time $t$ is then calculated using the transformations (\ref{mix3})
with the result
\begin{equation}\label{exnue4}
{\cal P}^r_{ee}(t-t^\prime, {\bar k})= \i 
 u^{r \dag}_{k,1}{\mathrm e}^{\i\omega_{k,1}(t-t')}\,
 \widetilde{G}^>_{ee}(t-t^\prime, {\bar k})\,\gamma^0 u^r_{k,1}
\;.\end{equation}
Here, $\widetilde{G}^>_{ee}(t-t^\prime, {\bar k})$ 
denotes the Fourier transform
of the unordered Green function (or Wightman function)
in the state that $\alpha_{k,e}^{r \dag}(t^\prime)$ is acting upon.
 The Fock space vacuum state we use is
$|0\rangle_{12}$, and the relative Wightman function is
\begin{equation}\label{gre0}
\i G^{> \alpha \beta}_{ee}(t,{\bar x};t',{\bar y})= {}_{1,2}\langle0| 
\nu^{\alpha}_{e}(t,{\bar x}) \;
\bar{\nu}^{\beta}_{e}(t',{\bar y})|0\rangle_{1,2} 
\;.\end{equation}
We obtain, for $t'=0$,
\begin{equation}\label{gre2}
\widetilde{G}^{>\,\alpha\beta}_{ee}(t,{\bar k}) =\\
 -\i\sum_{r} 
 \left( \cos^2\!\theta\;
 {\mathrm e}^{-\i\omega_{k,1} t}\; u^{r,\alpha}_{k,1}\;
  \bar{u}^{r,\beta}_{k,1} \; 
 + \sin^2\!\theta\;
 {\mathrm e}^{-\i\omega_{k,2} t} \;u^{r,\alpha}_{k,2} \;
 \bar{u}^{r,\beta}_{k,2}\;
 \right) 
\;.\end{equation}
The probability amplitude for the survival of the single neutrino
state from initial time $t^\prime=0$ up to time $t$ which follows
from eq. (\ref{exnue4}), is, independent of the spin orientation,
\begin{equation}
{\cal P}_{ee}(t,{\bar k})
  =\cos^2\!\theta\,
 + \sin^2\!\theta\,|U_k|^{2}\, {\mathrm e}^{-\i(\omega_{k,2}-
\omega_{k,1}) t}
\;.\end{equation}

The ``survival'' probability amplitude of an electronic neutrino state for 
very small times should give 1 if our definition (\ref{gre0}) 
were correct: $\lim_{t\rightarrow 0^+} {\cal
P}_{ee}(t) = 1$. One obtains instead
\begin{equation}\label{pee2}
{\cal P}_{ee}(0^+, {\bar k})=
\cos^2\!\theta + \sin^2\!\theta\, |U_k|^2 < 1
\;.\end{equation}
This means that the vacuum state $|0\rangle_{1,2}$, which contains a
pair condensate of the flavor neutrinos, cannot be used
in the calculation of Green's functions. We note that the above
result for the survival probability amplitude
reproduces the Pontecorvo oscillation formula in the
relativistic limit $|{\bar k}|>>\sqrt{m_{1}m_{2}}$
where $|U_k|^2 \simeq 1$.

At time $t = 0$  the state 
which
is annihilated by the flavor operators $\alpha_{e,\mu}(0)$ and
$\beta_{e,\mu}(0)$, is the ``flavor vacuum'': $|0\rangle_{e,\mu}\equiv
{G}^{-1}(\theta,0) |0\rangle_{1,2}$.
The correct definition of the Wightman function 
for $\nu_{e}$ is then
\begin{equation}\label{gfu1}
\i G^{>\alpha \beta}_{ee}(t,{\bar x};0,{\bar y}) = 
 {}_{e,\mu}\langle0| 
\nu^{\alpha}_{e}(t,{\bar x}) \; \bar{\nu}^{\beta}_{e}(0,{\bar y})
|0\rangle_{e,\mu}
\;.\end{equation}
to be compared with eq.(\ref{gre0}) where $|0\rangle_{1,2}$ is
used.
An important point in the definition (\ref{gfu1}) arises from
the fact that one may not directly compare
states at different times. Since the generator of the mixing 
transformations is time dependent (see Sec. 2), 
the relative orientation of the
$\nu_1$--$\nu_2$ Fock space and the flavor Fock space is twisted
with time. Therefore the comparison of states at different times
necessitates a {\em parallel transport\/} of these states to
a common point of reference. The definition (\ref{gfu1}) includes
this concept of parallel transport, and indeed the fixation of a zero
point is a gauge fixing also in the usual way of QFT:
the richness of
the geometric structure of mixing transformations, i.e., gauge fixing, 
parallel transport and geometric
phases will be considered in a separate
publication. 

Also, we note that up to a phase factor the same function as 
(\ref{gfu1}) is obtained 
if on {\em both sides\/} the flavor vacuum state is replaced by the state
$
|0_t\rangle_{e,\mu}  ={G}^{-1}(\theta,t) |0\rangle_{1,2}$. This is in 
agreement with our comment on the parallel transport: one may
compare the two state vectors at either end of the time trajectory.
However, a different result is obtained when replacing the 
flavor vacuum state on both sides
of eq. (\ref{gfu1}) by $|0_\tau\rangle_{e,\mu} 
={G}^{-1}(\theta,\tau) |0\rangle_{1,2}$ with arbitrary $\tau$.
The reason is that the product 
$\nu^{\dag}_e(t,{\bar x})|0_\tau\rangle_{e,\mu}$
cannot be interpreted as a single electron neutrino created at time $\tau$.

Eq. (\ref{gfu1}) is conveniently expressed in terms of 
anticommutators at different times
as
\begin{eqnarray}\nonumber
\widetilde{G}^{>\alpha\beta}_{ee}(t,{\bar k})& =& -\i\sum_{r}
\left[ 
 u^{r,\alpha}_{k,1}\,{\mathrm e}^{-\i\omega_{k,1}t}\; \bar{u}^{r,\beta }_{k,1}
 \,\left\{\alpha^r_{k,e}(t),\alpha^{r\dag}_{k,e}(0)\right\}
 \right.\\
\label{gfu2}
&&\;\;\;\left.+
 v^{r,\alpha}_{-k,1}\,{\mathrm e}^{\i\omega_{k,1}t}\; \bar{u}^{r,\beta}_{k,1}
 \,\left\{\beta^{r\dag}_{-k,e}(t),\alpha^{r\dag}_{k,e}(0)\right\}\right]
\;.\end{eqnarray}
This structure shows that
our definition of probability amplitudes singles out one anticommutator by 
time:
\begin{eqnarray}\nonumber
{\cal P}^r_{ee} (t,{\bar k})
 &=& \i u^{r \dag}_{k,1}{\mathrm e}^{\i\omega_{k,1}t}\,
 \widetilde{G}^>_{ee}(t,{\bar k})\,\gamma^0 u^r_{k,1} = 
\left\{\alpha^r_{k,e}(t),\alpha^{r\dag}_{k,e}(0)\right\}\\
 &=&  \cos^{2}\!\theta\,
 + \sin^2\!\theta\,\left[ |U_k|^{2} {\mathrm e}^{-\i
 (\omega_{k,2}-\omega_{k,1}) t}
     + |V_k|^{2} {\mathrm e}^{\i(\omega_{k,2}+\omega_{k,1}) t}\right]\\ 
\nonumber
{\cal P}^r_{\bar{e}e} (t,{\bar k})
 &=&\i v^{r \dag}_{-k,1}{\mathrm e}^{-\i\omega_{k,1}t}\,
 \widetilde{G}^>_{ee}(t,{\bar k})\,\gamma^0 u^r_{k,1} =
  \left\{\beta^{r\dag}_{-k,e}(t),\alpha^{r\dag}_{k,e}(0)\right\}\\
\label{pee4}
 &=&  \epsilon^r\,|U_k| |V_k|\,\sin^{2}\!\theta\,
 \left[ {\mathrm e}^{\i (\omega_{k,2}-\omega_{k,1})t}\;
-\;{\mathrm e}^{-\i (\omega_{k,2}+\omega_{k,1})t} \right]
\end{eqnarray}
Note that the probability amplitude is now normalized correctly:  
$lim_{t\rightarrow 0^+} {\cal P}_{ee}(t,{\bar k})=1$.
For completeness we
define also the ``mixed'' Green function
\begin{equation}
\i G^{>\alpha \beta}_{\mu e}(t,{\bar x};0,{\bar y}) =_{e,\mu}\langle0| 
  \nu^{\alpha}_{\mu}(t,{\bar x}) \; \bar{\nu}^{\beta}_{e}(0,{\bar y})
|0\rangle_{e,\mu}
\;.\end{equation}
\begin{eqnarray}\nonumber
{\cal P}^r_{\mu e}(t,{\bar k})
&=& \i u^{r \dag}_{k,2}{\mathrm e}^{\i\omega_{k,2}t}\,
 \widetilde{G}^>_{\mu e}(t,{\bar k})\,\gamma^0 u^r_{k,1} = 
\left\{\alpha^r_{k,\mu}(t),\alpha^{r\dag}_{k,e}(0)\right\} \\
\label{pee5}
&=&\;|U_k|\;\cos\!\theta\;\sin\!\theta \left[1\;-\;
 {\mathrm e}^{\i (\omega_{k,2}-\omega_{k,1}) t}\right]\\ 
\nonumber
{\cal P}^r_{\bar{\mu} e}(t,{\bar k})
&=&\i v^{r \dag}_{-k,2}{\mathrm e}^{-\i\omega_{k,2}t}\,
 \widetilde{G}^>_{\mu e}(t,{\bar k})\,\gamma^0 u^r_{k,1} = 
\left\{\beta^{r\dag}_{-k,\mu}(t),\alpha^{r\dag}_{k,e}(0)\right\} \\
\label{pee6}
&=&\;\epsilon^r\;|V_k|\;\cos\!\theta\;\sin\!\theta \left[1\;
-\;{\mathrm e}^{-\i (\omega_{k,2}+\omega_{k,1})t} \right]
\;.\end{eqnarray}
All 
other anticommutators with $\alpha^{\dag}_e(0)$
vanish. The conservation of the total probability then requires that
\begin{equation}
\left|{\cal P}_{ee}(t,{\bar k})\right|^2 + 
\left|{\cal P}_{\bar{e}e}(t,{\bar k})\right|^2 +
\left|{\cal P}_{\mu e}(t,{\bar k})\right|^2 +
\left|{\cal P}_{\bar{\mu}e}(t,{\bar k})\right|^2 =1
\,,\end{equation}
which is of course satisfied by our calculation.

Further work on the Green's functions formalism is in progress.

This work has been partially supported by EC contract CHRX-CT94-0423.

We are grateful to Mario Greco for giving us the opportunity to present 
this paper at Les Rencontre de Physique de la Vall\'ee d'Aoste, "Results
and Perspectives in Particle Physics", March 1996.



\newpage

\begin{center}
Table I:
The values of $\sqrt{ m_{1} m_{2} } 
$ and of $a$ for given values of $m_{1}$ and $m_{2}$.
\end{center}  
\begin{center}
\begin{tabular}{|c||c|c|c|c|} \hline\hline
&{\em $m_{1}$(eV)} & {\em $m_{2}$(KeV)} & {\em $\sqrt{m_{1}m_{2}}$(KeV)} &
{\em $a$}  \\ \hline
$A$ & $ 5   $   &  $ 250 $    &  $   1.12    $    &  $ \sim 5 \cdot 10^{4}  $ 
\\ \hline
$B$ & $ 2.5   $   &  $ 250 $    &  $   0.79    $    &  $ \sim 1 \cdot 10^{5}  $
\\ \hline
$C$ & $ 5   $   &  $ 200 $    &  $   1    $    &  $ \sim 4 \cdot 10^{4}  $ 
\\ \hline
$D$ & $  1  $   &  $ 100 $    &  $  0.32  $  &  $ \sim 1 \cdot 10^{5}  $  \\
\hline
$E$ & $ 0.5 $   &  $ 50  $    &  $  0.15  $  &  $ \sim 1 \cdot 10^{5}  $  \\
\hline
$F$ & $ 0.5 $   &  $ 1   $    &  $  0.02  $  &  $ \sim 2 \cdot 10^{3}  $  \\
\hline
\hline
\end{tabular}
\end{center}

\bigskip
\medskip

\begin{center}
Table II: $|U(p_{\epsilon},a)|^{2} \; $vs.$ \;k_{\epsilon}$.
\end{center} 
\begin{center}
\begin{tabular}{|c||c|c||c|c||c|c|} \hline\hline
 
& {\em $|U(1,a)|^{2}$}           &{\em $k_{1}$(KeV)} 
& {\em $|U(p_{1/2},a)|^{2}$}     & {\em $k_{1/2}$(KeV)}  
& {\em $|U(p_{1/10},a)|^{2}$}    & {\em $k_{1/10}$(KeV)} \\ \hline
$A$ & $ \simeq 0.5 $  &  $ 1.12 $    &  $ \simeq 0.75 $  &  $ \simeq 146 $
&  $ \simeq 0.95 $  &  $ \simeq 519 $ \\ \hline
$B$ & $ \simeq 0.5 $  &  $ 0.79 $    &  $ \simeq 0.75 $  &  $ \simeq 145 $
&  $ \simeq 0.95 $  &  $ \simeq 518 $ \\ \hline
$C$ & $ \simeq 0.5 $  &  $ 1 $    &  $ \simeq 0.75 $  &  $ \simeq 117 $
&  $ \simeq 0.95 $  &  $ \simeq 415 $ \\ \hline
$D$ & $ \simeq 0.5 $  &  $ 0.32 $  &  $ \simeq 0.75 $  &  $ \simeq  58 $
&  $ \simeq 0.95 $  &  $ \simeq 206 $ \\ \hline
$E$ & $ \simeq 0.5 $  &  $ 0.16 $  &  $ \simeq 0.75 $  &  $ \simeq  29 $
&  $ \simeq 0.95 $  &  $ \simeq 104 $ \\ \hline
$F$ & $ \simeq 0.5 $  &  $ 0.02 $  &  $ \simeq 0.75 $  &  $ \simeq 0.6 $
&  $ \simeq 0.95 $  &  $ \simeq  2  $ \\ \hline
\hline
\end{tabular}
\end{center}

\bigskip
\medskip

\begin{center}
Table III: $|U(p_{\epsilon},a)|^{2} \;$ vs.$ \;k_{\epsilon}$ for
$m_{1}\simeq 0$ and different values of $m_{2}$.
\end{center} 
\begin{center}
\begin{tabular}{|c|c||c|c||c|c|} \hline\hline
{\em $m_{1}$(eV) }              &{\em $m_{2}$(KeV)} 
& {\em $|U(p_{1/2},a)|^{2}$}     & {\em $k_{1/2}$(KeV)}  
& {\em $|U(p_{1/10},a)|^{2}$}    & {\em $k_{1/10}$(KeV)} \\ \hline
$ \simeq 0 $  &  $ 250 $  &  $ \simeq 0.75 $  &  $ \simeq 144 $
&  $ \simeq 0.95 $  &  $ \simeq 516 $ \\ \hline
$ \simeq 0 $  &  $ 200 $  &  $ \simeq 0.75 $  &  $ \simeq 115  $
&  $ \simeq 0.95 $  &  $ \simeq 413 $ \\ \hline
$ \simeq 0 $  &  $ 100 $   &  $ \simeq 0.75 $  &  $ \simeq 57 $
&  $ \simeq 0.95 $  &  $ \simeq 206 $ \\ \hline
$ \simeq 0 $  &  $ 50 $  &  $ \simeq 0.75 $  &  $ \simeq 29  $
&  $ \simeq 0.95 $  &  $ \simeq 103 $ \\ \hline
\hline
\end{tabular}
\end{center}

\end{document}